\documentclass[prl,amsmath,twocolumn,showpacs,superscriptaddress]{revtex4}
\usepackage{graphicx}
\begin{document}

\title{ 
Non-normalizable densities in strong anomalous diffusion: beyond 
the central limit theorem
 }
\author{Adi Rebenshtok}
\affiliation{Department of Physics, Institute of Nanotechnology and Advanced Materials, Bar-Ilan University, Ramat-Gan,
52900, Israel}
\author{Sergey Denisov}
\affiliation{Sumy State University, Rimsky-Korsakov Street 2, 40007 Sumy, Ukraine}
\affiliation{Institute  of Physics, University of Augsburg,
Universit\"atsstrasse 1, D-86135,  Augsburg 
Germany}
\author{Peter H\"anggi}
\affiliation{Institute  of Physics, University of Augsburg,
Universit\"atsstrasse 1, D-86135,  Augsburg 
Germany}
\author{Eli Barkai}
\affiliation{Department of Physics, Institute of Nanotechnology and Advanced Materials, Bar-Ilan University, Ramat-Gan, 
52900, Israel}

\begin{abstract}
Strong anomalous diffusion, where $\langle |x(t)|^q \rangle \sim t^{q \nu(q)}$
with a nonlinear spectrum $\nu(q) \neq \mbox{const}$, is wide spread 
and has been  found in various  nonlinear dynamical systems 
and
experiments on active transport in living cells. Using a stochastic approach
we  show how this phenomena
is related to infinite covariant densities, i.e., the asymptotic states
of these  systems
are described by non-normalizable distribution functions.
 Our work shows that the concept
of  infinite covariant densities
plays an important role in the statistical description of open systems
exhibiting multi-fractal anomalous diffusion, as it is complementary to 
the central limit theorem.  
\end{abstract}
\maketitle


 Consider particles diffusing in a medium 
with  their total number  
conserved.
Within a probabilistic approach the density $P(x,t)$ is normalized
to unity 
$\int_{-\infty} ^\infty P(x,t) {\rm d}x =1$
for any time $t$.
It follows naturally
that the steady state of a system in equilibrium
is normalized, e.g., the Boltzmann-Gibbs distribution
which serves as the basis of statistical physics
in thermal equilibrium. 
In contrast, infinite ergodic theory is a branch of mathematics 
that investigates
dynamical systems whose invariant density
is non-normalizable \cite{Aaronson,Thaler}.  
Current  models characterized by non-normalized densities include
intermittent maps \cite{KorabelPesin,KorabelINF,Akimotoprl} 
and the  momentum distribution
of particles in optical lattice \cite{KesslerPRL,LutzNature,Holz}. 
These systems attain equilibrium, i.e., the non-normalized densities
is an extension of Boltzmann-Gibbs like states \cite{KesslerPRL,LutzNature}, 
 and
unfortunately we conclude that real-life applications 
 of infinite ergodic theory are limited. 
Here we investigate unbounded systems which
are not in equilibrium, inspired by infinite ergodic theory,
 we find that non-normalized covariant densities (defined below)
play an important
role in the description of anomalous diffusion.
The potential applications of infinite covariant densities
in real world experiments
is shown to be vast. 
These uncommon densities provide a detailed description of 
the rare fluctuations in  strong anomalous diffusion. 

 Strong 
anomalous diffusion deals with processes, 
with a long time $t$ asymptotic behavior, satisfying 
%
$\langle |x(t)|^q \rangle \sim t ^{q \nu(q)}$,
%
$q>0$, where $\nu(q)$ is not a constant \cite{Strong}, in contrast 
 with standard Brownian motion where $\nu(q) =1/2$.
Such diffusion has been detected in a variety of processes: 
in the transport
in  two-dimensional incompressible velocity fields \cite{Strong},  
particle spreading
in billiard systems  
\cite{Edelman,Artuso,Armstead,Sanders,Courbage},
avalanche dynamics of  sand pile models \cite{Sandpile}, 
and in statistics of occupation times of renewal processes \cite{GL}.
 Recent  experiments on the active transport of
polystyrene  beads in living cells
\cite{Naama},
theoretical investigation of
cold  atoms in optical lattices \cite{DechantPRL}   
and 
flows in porous media \cite{Anna}
further confirmed the generality of strong anomalous diffusion. 
Remarkably, in all these systems a piecewise-linear scaling, 
with  $q\nu(q) \sim q$ 
below some critical value of $q$, and with  $q\nu(q)= q - b$ above this
critical value, was found. 
An analytical approach to this 
bilinear scaling, for deterministic
chaotic
dynamics was presented in Ref. \cite{Artuso}. 
Such strong anomalous diffusion is an indication of the breakdown of 
mono-scaling theories
which predict 
$P(x,t) \sim t^{-\nu} f(x/t^{\nu})$,
 e.g., normal diffusion where
$f(\cdot)$ is Gaussian.  
In this Letter we explain  how this breakdown 
is related to non-normalizable densities. By investigating a large class
of stochastic processes, the so-called L\'evy walks
\cite{Shles,Klafter,Bouchaud,Review},
we demonstrate how these densities describe
statistics of strong anomalous diffusion. Our work shows
how  infinite covariant densities are complementary to the
L\'evy-Gauss central limit theorem, which presents the mathematical foundation
of diffusion phenomena.    

{\em The  L\'evy walk model} 
\cite{Shles,Klafter,Bouchaud,Review}
 is a widely applicable process
describing  
strong anomalous diffusion \cite{Andersen}.
To demonstrate the broad validity of our approach
 we consider two different classes of the model and four examples.
In the {\em velocity model} \cite{ZK}, 
 a particle in one dimension starts on the origin
$x=0$ at time $t=0$
and  travels with velocity
$v_1$, drawn from a probability density function (PDF) $F(v)$.
 The duration
of the traveling event $\tau_1$ is
 drawn from the PDF $\psi(\tau)$.
 The process is then renewed,
namely, a new velocity $v_2$  and flight duration
$\tau_2$ are drawn from $F(v)$ and $\psi(\tau)$ respectively. 
The process continues
in this manner until time $t$. The position of the particle is
 $x = \int_0 ^t v(t) {\rm d} t$, as usual.  In the {\em jump model}
 \cite{ZK,KBS},
a particle
first waits on  $x=0$ for time $\tau$, drawn from $\psi(\tau)$,
 and performs a  jump
with probability $1/2$ to a distance $x=v_0\tau$ or $x=-v_0\tau$, where
$v_0=\mbox{const}$. 
The process is then renewed. 
For the velocity model we assume that $F(v) = F(-v)$
 and that all the moments of $F(v)$ are finite.
 We will address three cases:  
(i) a two state velocity model,
$F(v)= \left[ \delta(v-v_0) + \delta(v + v_0) \right]/2$,
 (ii) a Gaussian and (iii) exponential  velocity PDFs.
The main ingredient of the L\'evy walk
 model is the
waiting time PDF
of the power law form
\begin{equation}
\psi(\tau)  \sim  {A \over |\Gamma(-\alpha)|}  \tau^{- (1 + \alpha)} 
\label{eqpsi}
\end{equation}
with $1<\alpha<2$ and $A>0$. This choice of parameters insures that the average
waiting time, $\langle \tau \rangle$,  is finite but
the variance of the waiting time is infinite. 
These types of random walks have been widely investigated  and 
specific values of $\alpha$ have been recorded in several experiments 
\cite{Ott,Solomon,Hoog,Wiersma,Sims,Jager,Harris,Sagi}
and calculated  from first principle models \cite{KesBarPRL,Zabu}
. 

{\em Montroll-Weiss equation.}
Let $P(x,t)$ be the PDF of the particle's position 
at time $t$. The well
known Montroll-Weiss equation \cite{ZK,KBS}
gives the Fourier-Laplace transform of $P(x,t)$ 
for the velocity model \cite{Carry}
\begin{equation}
P(k,u) = { \overline{W}(k,u) \over 1 - \overline{\psi}(k,u) },
\label{eqMW}
\end{equation}
where 
%
$\overline{\psi}(k,u) =  \int_0 ^\infty \int_{-\infty} ^\infty  \exp(- u \tau + i k v \tau ) F (v) \psi(\tau) {\rm d} \tau {\rm d} v$
and similarly for $\overline{W}(k,u)$ with 
 $W(t) = \int_t ^\infty \psi(\tau) {\rm d} \tau$.
Here $W(t)$ is the persistence 
probability to reach $x$ in a single travelling event.
%
Henceforth we use the convention that the variables in
a function's  parenthesis, e.g., $P(x,t)$ or $P(k,u)$,
define the space we are working in. 

\begin{figure}\begin{center}
\includegraphics[width=0.5\textwidth]{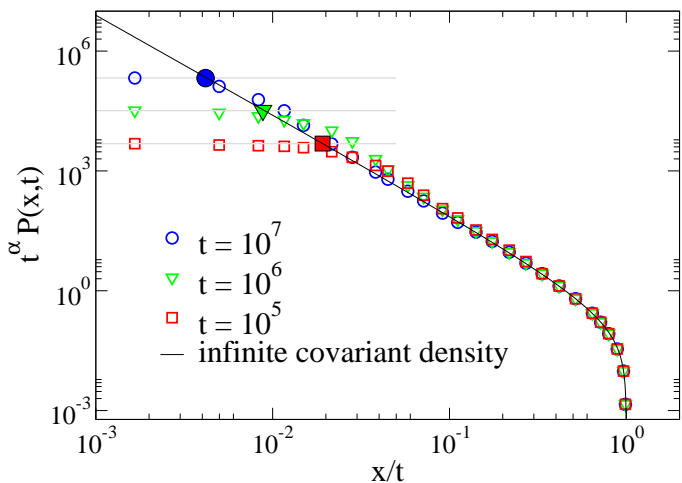}
\end{center}
\caption{ (color online)
Rescaled PDFs
 $t^\alpha P(x,t)$ (open symbols)  versus $x/t$ for the jump model
with $\alpha=3/2,A=v_0=1$ \cite{remarkSim}
 for three different times.
The bold black line depicts the  ICD of the process
Eq. (\ref{eqjump}).
Error bars  are smaller then the size of the symbols.
Big filled symbols indicate the locations of the crossover velocities
$\overline{v}_c$ and horizontal lines are L\'evy's central limit theory 
for
$t^\alpha P(x=0,t)$
(see the main text for more details). 
}
\label{fig1}
\end{figure}

The Fourier transform $P(k,t)$ is Taylor expanded 
\begin{equation}
P(k,t) = 1 + \sum_{n=1} ^\infty { (i k)^{ n} \langle x^{n}(t) \rangle \over
n ! }, 
\label{eq03}
\end{equation}
where the first term is the normalization $P(k=0,t)=1$. 
Using Eq. (\ref{eqMW})  we obtain the integer  moments of the process 
in the long time limit. The $n$-th moment is obtained by differentiating
$\langle x^n(u) \rangle = d^n P(k,u) /d (ik)^n|_{k=0}$ 
with 
 $\psi(u) \sim 1 - \langle \tau \rangle u + A u^\alpha$ 
and then by Laplace inversion.
This is a standard procedure  for $n=2$ \cite{Review} 
and  we use the  Fa\'a di Bruno formula \cite{Warren}
to obtain the exact asymptotic expressions
for all the higher order moments.
The even  $n$-th moment of the two state velocity model
is
\begin{equation}
\langle x^{ n}(t) \rangle \sim 
{  n \over ( n - \alpha) ( n + 1 - \alpha) } { A \over |\Gamma(1-\alpha)| \langle \tau\rangle} (v_0)^{  n} t^{  n + 1 - \alpha}, 
\label{eq04}
\end{equation}
while odd moments are zero,
due to the symmetry of $F(v)$.  
The process exhibits super-diffusion $\langle x^2 \rangle\sim t^{3 - \alpha}$
because $1<\alpha<2$. 
We insert Eq. (\ref{eq04}) in Eq. (\ref{eq03}) and find
\begin{equation}
P_A(k,t) \sim 1 + t^{1 -\alpha} { A \over |\Gamma(1 -\alpha)| \langle \tau \rangle}  \tilde{f}_\alpha(i k v_0 t), 
\label{eq05}
\end{equation}
where the subscript $A$ in $P_A(k,t)$ stands  for the $t\to \infty$
 asymptotics and 
\begin{equation}
\tilde{f}_\alpha ( i y ) = \sum_{n=1}^\infty { (i y)^{2 n} \over (2 n -1)! (2n -\alpha) (2n + 1 - \alpha) }.
\label{eq06}
\end{equation}
Summing this series
 we get 
\begin{widetext}
\begin{equation}
\tilde{f}_\alpha( i y) = y^2 \left[
 { 1 \over 3- \alpha}\ _1F_2 \left( { 3 - \alpha \over 2} ; {3 \over 2} , {5 - \alpha \over 2} ; { - y^2 \over 4} \right) 
 -{1 \over 2 - \alpha}\ _1F_2 \left( 1 - {\alpha \over 2}; { 3 \over 2} , 2 - {\alpha \over 2} ; { - y^2 \over 4} \right)
\right] 
\label{eq07}
\end{equation}
where $_1 F_2$ is a Hypergeometric function. 
Next we invert the asymptotic expression $P_{A}(k,t)$, Eq. (\ref{eq05}),
 using the inverse Fourier transform and Eq. (\ref{eq07}).
Since we use the exact expressions for the
long time behavior of the  moments of the process, one may be
tempted to  
believe that this procedure finally yields the long time limit 
of the normalized spreading packet $P(x,t)$.
 Indeed for normal transport,
for  example, when the waiting times are exponentially distributed,
this procedure yields the familiar  Gaussian distribution. 
Instead, for the two state L\'evy walk model, we find
\begin{equation}
P_A(x,t)= 
 {1 \over t^\alpha} { A  \over 2  v_0 \langle \tau \rangle \left|\Gamma(1-\alpha)\right|}   \left|{x \over v_0 t}\right|^{-(1+ \alpha)} \left[ \alpha - \left( \alpha - 1 \right) \left|{x \over v_0 t}\right| \right] \ \mbox{for} 
 \ \    |x|<v_0 t, \ \ x\neq 0,
\label{eq08}
\end{equation}
\end{widetext}
while $P_A(x,t)=0$ for $|x|> v_0 t$. 
Notice that in the vicinity of $x\to 0$ we find $P_A(x,t) \sim |x|^{-1 - \alpha}$,
 therefore it is non-normalizable.  
Within our approach we first take the long time limit
(in the 
 calculation of the moments) and only then perform the inverse Fourier
transform. These two mathematical operations do not commute,
namely, the inverse Fourier transform of $P(k,t)$, for any finite
$t$, yields a normalized
density. 
Still as we proceed
to show,  the non-normalized state
Eq. (\ref{eq08}) describes statistical properties of
the spreading packet of particles and hence physical reality.

{\em Infinite covariant density.} We now define the infinite  
covariant density (ICD) according to
\begin{equation}
\lim_{t\to \infty} t^\alpha P(x,t) = I_{\rm{cd}} (\overline{v})
\label{eq09}
\end{equation}
where $\overline{v} \equiv { x/ t}  
=\int_0 ^t  v(t) {\rm d} t/t$ 
is the $t\to \infty$ time averaged velocity of the particle.
Since both $P(x,t)$ and $P_A(x,t)$  provide the asymptotic
 moments of the process,
in the asymptotic regime
Eqs. (\ref{eq08},\ref{eq09}) give 
\begin{equation}
I_{\rm{cd}} (\overline{v}) = K_\alpha c_\alpha |\overline{v}|^{-(1+\alpha)} \left[ 1 - { \alpha -1 \over \alpha} {|\overline{v}| \over v_0} \right],
\label{eq10}
\end{equation}
if $|\overline{v}|/v_0 \le 1$, otherwise $I_{\rm{cd}}(\overline{v})=0$, 
and $c_\alpha = \sin (\pi \alpha/2) \Gamma(1 + \alpha) / \pi$,
$K_\alpha = A \langle |v|^\alpha \rangle |\cos(\pi \alpha 
/2)|/\langle \tau \rangle$ with $\langle |v|^\alpha \rangle= \int_{-\infty} ^\infty |v|^\alpha F(v) {\rm d} v$. 
Here $K_\alpha$ is the anomalous diffusion constant, a measurable observable, 
soon to be discussed. 

\begin{figure}\begin{center}
\includegraphics[width=0.5\textwidth]{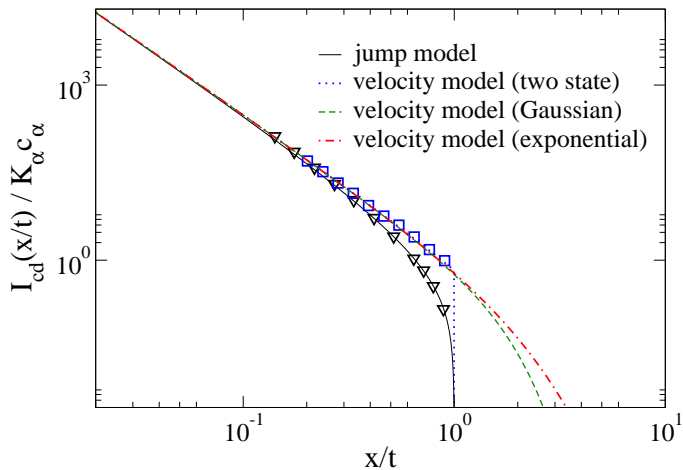}
\end{center}
\caption{
(color online) ICDs for different L\'evy walk models,
with $\alpha=3/2$. 
 All densities exhibit 
characteristic non-integrability on $x/t\to 0$ and collapse for small $x/t$
on a master curve Eq. 
(\ref{eq22}).
Open blue squares (velocity model) and black triangles 
(jump model) are 
simulation results for $t=10^6$.
}
\label{fig2}
\end{figure}

{\em Universality.} 
The non-normalized  density is a generic
feature of L\'evy walk processes. A detailed calculation for 
the jump model yields
\begin{equation}
 I_{\rm{cd}} ^{\rm{jump}} (\overline{v}) = K_\alpha c_\alpha |\overline{v}|^{-(1+\alpha)} \left[ 1 - |\overline{v}/v_0|\right]
\label{eqjump}
\end{equation}
for $|\overline{v}/v_0|<1$ and 
 $K_\alpha=A (v_0)^\alpha |\cos(\pi \alpha /2)|/\langle \tau \rangle$ \cite{pathways}. 
 The ICDs
  Eqs. (\ref{eq10},\ref{eqjump})
exhibit universal features:  the scaling variable is
$\overline{v} = x/t$, the power law divergence in the limit
 $\overline{v} \to 0$,
and   the dependence of the density only on asymptotic
properties of $\psi(\tau)$. 
Notice that for the jump model, Eq. (\ref{eqjump}) shows
that we obtain zero density at $|\overline{v}|= v_0$, while
for the two-state velocity model Eq. (\ref{eq10}) shows a finite density at
$\overline{v}=v_0$. This is because the velocity 
mechanism propagates the particles further if compared with the jump approach.
 This indicates that the ICD
can distinguish between 
these two models and hence is
a valuable tool in data analysis \cite{Soko}.

 For the Gaussian  velocity distribution, 
$F(v) = (\sqrt{2 \pi } v_0)^{-1}  \exp(-v^2/2 v_0^2)$,
we find
\begin{widetext}
\begin{equation}
I_{\rm{cd}} ^{{\rm G}} (\overline{v}) =K_\alpha c_\alpha  | \bar{v} |^{-(1+\alpha)} \Bigg(1 -\frac{2^{\frac{1}{2}}\Gamma \big(\frac{\alpha}{2}\big)}{ \alpha \Gamma \big(\frac{\alpha-1}{2}\big) } \frac{|\bar{v}|}{v_0}  \Bigg)
- \frac{K_\alpha C_\alpha}{2^{\frac{1+\alpha}{2}} \alpha^2 \Gamma(\frac{3+\alpha}{2})v_0^{1+\alpha}  } {}_3 F _3 \bigg(\frac{3}{2},\frac{\alpha}{2},\frac{1+\alpha}{2};1+\frac{\alpha}{2},\frac{3+\alpha}{2},\frac{1}{2};-\frac{\bar{v}^2}{2 v_0^2}\bigg),
\label{eqlong}
\end{equation}
\end{widetext}
which once more exhibits the characteristic divergence at $\overline{v}\to 0$.
We also obtained the ICD for an exponential
distribution of velocities 
$F(v) =  (\sqrt{2} v_0)^{-1} \exp( -\sqrt{2} |v|/v_0)$, 
which is  soon 
presented graphically 
together with the ICDs
of the  other models.

{\em Significance  of  the ICD  
and its physical meaning.} Clearly the ICD
is not renormalizable due to its behavior close to $\overline{v} \to 0$.
Hence it is not immediately obvious that it describes the concentration
of particles.   
The role of the ICD is however two-fold. 
 First, we define observables which are integrable
with respect to the ICD, e.g., $|\overline{v}|^{q}$
provided that $q>\alpha$. The averages of these observables
are given by the ICD, for example,
\begin{equation}
\langle |\overline{v}|^q \rangle = t^{1-\alpha} \int_{-\infty} ^\infty |\overline{v}|^q I_{\rm{cd}}(\overline{v}) {\rm d} \overline{v}
\label{eq10aa}
\end{equation}
so that $\langle |x(t)|^{q} \rangle = t^q \langle
|\overline{v}|^q \rangle$ \cite{remarkP}.
Second, to attain the ICD from data, one should plot 
$t^\alpha P(x,t)$ versus $x/t$, where here $P(x,t)$ is the concentration
of particles. This type of plot will collapse in
the limit of long times onto a master curve: the ICD,
Eq. (\ref{eq09}). Thus, the ICD
is a property of the spreading packet, and not merely a
mathematical tool
with which we calculate averages. 
This procedure is presented in Fig. \ref{fig1}, where
we see excellent agreement between the theory and simulations.

{\em Relation between ICD and the anomalous diffusion
constant $K_\alpha$}. From Fig. \ref{fig1},  we see that the convergence
to the ICD
is slow in  the vicinity of the origin 
$\overline{v}=x/t\to 0$. 
In this region the packet $P(x,t)$  satisfies  the 
fractional diffusion equation \cite{Review}
\begin{equation}
{\partial P_{\rm{cen}}(x,t) \over \partial t} = K_\alpha \nabla^\alpha P_{\rm{cen}}(x,t),
\label{eq11}
\end{equation}
where the fractional derivative is given by its Fourier representation
$\nabla^\alpha \to -|k|^\alpha$. 
This equation is derived from 
 Eq.  
(\ref{eqMW}), 
 assuming  
$x\sim
t^{1/\alpha}$, while the ICD implies $x\sim t$. 
The  solution of Eq. (\ref{eq11}) is 
\begin{equation}
P_{\rm{cen}} (x,t) \sim { 1 \over (K_\alpha t)^{1/\alpha}} L_\alpha \left[ {  x \over \left(K_\alpha t\right)^{1/\alpha}}  \right],
\label{eq12}
\end{equation}
where $L_\alpha(x)$ is the symmetric  L\'evy distribution \cite{levy}.
 An observable like $|x|^q$ with
$q>\alpha$  is not integrable
with respect to the L\'evy PDF Eq. (\ref{eq12}). To see this
recall that $L_\alpha (x) \sim |x|^{- (1 +\alpha)}$ so 
the L\'evy PDF has a diverging second moment.
 Precisely for that
reason we need the ICD. Averages of  observables 
like $|x|^q$ with $q>\alpha$ are given by  the ICD as
mentioned,  while
the L\'evy PDF describes the scaling behavior 
of $P(x,t)$ at the central region only, and thus
cannot be used to give 
information even on the mean square displacement $\langle x^2(t) \rangle$.
 The opposite is also
true. For example, $|x|^q$ is non-integrable with respect
to the ICD for $q<\alpha$, and hence its mean should be
evaluated 
by using the L\'evy PDF, Eq. (\ref{eq12}). 
Thus L\'evy's central limit theorem \cite{levy}
 and the ICD provide complementary
information on the process, and both are needed for a
complete  statistical description
of the process.   

{\em Matching the two solutions.}
There must be  a connection between the L\'evy PDF and the ICD,
 since they
must match in intermediate regions. 
Using $L_\alpha(x) \sim c_\alpha |x|^{ -(1 + \alpha)}$ for large $|x|$
we obtain from Eq. 
(\ref{eq12})
\begin{equation}
t^\alpha P_{\rm{cen}} (x,t) \sim c_\alpha K_\alpha |x/t|^{-1 - \alpha},
\label{eq22}
\end{equation}
which is the same as that found in Eqs. 
(\ref{eq10},\ref{eqjump},\ref{eqlong}) 
when $\overline{v}\to 0$.
Hence the large $x$-behavior of $P_{\rm{cen}}(x,t)$ matches
the small argument behavior of the ICD. 
This, in turn, implies that if we measure the diffusion constant
$K_\alpha$ and exponent $\alpha$ by observing the central part of the 
packet, we can predict the behavior of the ICD at the origin
since
\begin{equation}
I_{\rm{cd}}(\overline{v}) \sim K_\alpha c_\alpha |\overline{v}|^{-(1+\alpha)} \ \mbox{for} \ \overline{v}\to 0.
\label{eq22}
\end{equation}
As shown in Fig. \ref{fig2},
this relation is universal for the class of models under investigation.

{\em Crossover behavior.} As shown,
the ICD is reached in the limit of 
infinite time. For finite but long  times there exists a crossover 
velocity $\overline{v}_c$ above which the ICD serves 
as a good approximation
for the density of particle, when scaled properly.
 At the origin we have
$t^\alpha P(x,t)|_{x=0} \sim t^\alpha L_{\alpha} (0) / (K_\alpha t)^{1/\alpha}$,
which is plotted in Fig. \ref{fig1} together with numerics.
 Using $t^\alpha P_{\rm{cen}}(x,t)|_{x=0} =
 I_{\rm{cd}} (\overline{v}_c)$ as the definition of the crossover
velocity $\overline{v}_c$ we find using Eqs. 
(\ref{eq12},\ref{eq22})
$\overline{v}_c = t^{ -(\alpha -1)/\alpha} (K_\alpha)^{1/\alpha} [c_\alpha / L_\alpha(0)] ^{1/(1+\alpha)}$ 
with $L_\alpha(0) =\Gamma(1+\alpha^{-1}) /\pi$.
This crossover velocity $\overline{v}_c$ is shown in Fig. \ref{fig1} (for three
times).
Since $\overline{v}_c$ approaches zero as a power law, the convergence
of numerical data to the
ICD is slow, especially when $\alpha \to 1$ from above \cite{remark2}.

{\em Strong anomalous diffusion.}
 Our results show a deep connection between strong anomalous 
diffusion and ICDs. Using Eqs.
(\ref{eq09},\ref{eq10aa},\ref{eq12}) we obtain
\begin{equation}
\langle |x(t)|^q \rangle = \left\{
\begin{array}{c c}
M_q ^{<}\  t^{q/\alpha}; & q < \alpha, \\
\ & \ \\
M_q ^{>}\ t^{q + 1 - \alpha }; &  q> \alpha.
\end{array}
\right.
\label{eqMom}
\end{equation}
Thus, the model exhibits strong anomalous diffusion similar to that found
in many systems briefly discussed in the introduction.
 The amplitudes $M_q ^{<}$
and $M_q ^{>}$ can be obtained  by using  L\'evy and ICDs 
respectively.
For the two state model we get
\begin{equation}
\begin{array}{c}
M_{q} ^< = (K_\alpha)^{q/\alpha} { \Gamma(1- q/\alpha) \over \Gamma(1 - q) \cos { \pi q \over 2} } \\
\  \\
M_{q} ^{>} = { 2 K_\alpha c_\alpha q (v_0)^{q-\alpha} \over \alpha
\left( q - \alpha \right) \left( q - \alpha + 1 \right) } .
\end{array}
\label{eqmm}
\end{equation} 
These amplitudes diverge as $q$ approaches $\alpha$ from below
or above, an indication of a dynamical phase transition.

{\em ICDs and Gaussian diffusion.} 
What happens to the ICD when the models yield
Gaussian diffusion,
i.e., when the variance of the waiting time PDF Eq.
(\ref{eqpsi})
is finite?  
In that case, the center part of the
packet  is described well by the Gaussian central limit theorem
\cite{ZK,KBS}. Still, 
 the outer parts of $P(x,t)$ are given by the ICD. 
For example, for $2<\alpha<3$ in Eq. (\ref{eqpsi}),  integer 
moments greater than the second are described by the ICD,
not by Gaussian statistics (details to be published). 
Thus, an ICD can be present even when the observed
diffusion at the central part of $P(x,t)$ is normal.

{\em Discussion and Summary.}
The fact that strong anomalous diffusion 
is a phenomenon observed in a great variety of different systems,
serves as evidence of the universality of the  ICD
concept. In addition, in many systems
where strong anomaly was found, 
one may pinpoint the signatures of the L\'evy walk type of dynamics. 
For example, the infinite horizon 
Lorentz model, with a tracer particle 
moving in an array of
scatterers,   
is known to induce long ballistic flights reminiscent
of the L\'evy walk model 
\cite{Bouchaud,Bou85,Zacharel}, with $\alpha \to 2$.
Likewise, the experiments on active transport in a live cell \cite{Naama}, 
where  the connection
to L\'evy walk was proven 
 by removing long jumps from
the trajectories and observing a transition from strongly anomalous to 
normal diffusion. 
Simply said, the applications of L\'evy walks are vast and it follows
that  the same is true for
the applications of ICDs.
 Beyond the conceptual beauty of non-normalizable densities
 which give rise to a certain universality, in the sense
of universal scaling $x\sim t$, the
scaling function itself captures some of the fine details of the model
(as opposed to L\'evy and Gaussian densities) 
and hence the information in the ICD
is crucial for precise characterisation of an anomalous process. 
 Related deviations from mono-scaling were also 
observed in models with a  drift \cite{Burioni},
though further work is called for
to relate ICDs to biased diffusion models. 


%
%

\acknowledgments
This  work  was supported by the  Israel Science  Foundation (AR and EB),
and the German Excellence 
Initiative ``Nanosystems Initiative Munich''(SD and PH).
EB thanks the Alexander von Humboldt foundation for its support.


\begin{thebibliography}{99}
\bibitem{pathways} With the $I_{\rm{cd}}(\overline{v})$ one may easily obtain the moments of the process which where previously calculated in:
 E. Barkai, {\em Chem. Phys.} {\bf 284}, 13 (2002).
\bibitem{remarkP}
 To derive Eq. (\ref{eq10aa}),
we separate the
$x$ integration domain, into two, using $P(x,t)\sim I_{\rm{cd}}(\overline{v})/t^\alpha$ 
to describe the large $x$ behavior, while the L\'evy density (soon to be discussed)
 describes
the small $x$ regime, the latter
in the $t \to \infty$  limit does not contribute to the integration 
when $q>\alpha$
(see also paragraph crossover behavior here
  and our future
publication). A key point is that
$\overline{v}^q$ compensates the divergence of the infinite
density on the origin, in such a way that the integral Eq.
(\ref{eq10aa}) is finite.
\bibitem{remarkSim}
The distribution 
for $t = 10^7$ has been sampled with $N = 10^{10}$ realizations.
The corresponding simulations 
were performed on two GPU clusters 
(each consisting of six $\rm{TESLA~K20XM}$ cards)
and took $200$ hours.
\end{thebibliography}
\end{document}